\documentclass[letterpaper, 10 pt, conference]{ieeeconf}  
\usepackage{graphicx}
\usepackage{subcaption}
\usepackage{amsmath}
\usepackage[normalem]{ulem}
\usepackage{multirow}
\usepackage{amsfonts}

\IEEEoverridecommandlockouts                              

\title{\LARGE \bf
Co-learning Single-Step Diffusion Upsampler and Downsampler with Two Discriminators and Distillation}

\author{Sohwi Kim$^{1}$, Tae-Kyun Kim$^{1}$ \\
$^{1}$School of Computing, KAIST
}

\begin{document}

\maketitle
\thispagestyle{empty}
\pagestyle{empty}

\begin{abstract}
Super-resolution (SR) aims to reconstruct high-resolution (HR) images from their low-resolution (LR) counterparts, often relying on effective downsampling to generate diverse and realistic training pairs. In this work, we propose a co-learning framework that jointly optimizes a single-step diffusion-based upsampler and a learnable downsampler, enhanced by two discriminators and a cyclic distillation strategy. Our learnable downsampler is designed to better capture realistic degradation patterns while preserving structural details in the LR domain, which is crucial for enhancing SR performance. By leveraging a diffusion-based approach, our model generates diverse LR-HR pairs during training, enabling robust learning across varying degradations. We demonstrate the effectiveness of our method on both general real-world and domain-specific face SR tasks, achieving state-of-the-art performance in both fidelity and perceptual quality. Our approach not only improves efficiency with a single inference step but also ensures high-quality image reconstruction, bridging the gap between synthetic and real-world SR scenarios.
\end{abstract}


\section{INTRODUCTION}
\label{sec:intro}

Real-world image quality often deteriorates due to various factors such as blurs, compression artifacts, color inaccuracies, and sensor noise. A major challenge in Super-Resolution (SR) is handling these unknown and complex degradation patterns. Traditional SR approaches generally assume simpler degradation models, such as Gaussian noise or bicubic downsampling. However, real-world degradation patterns are often much more complex. More advanced methods~\cite{degradation2, Wang2021, Zhang2018b} attempt to reflect these conditions more accurately but often struggle to generalize to unseen degradations. As a result, achieving realistic image reconstruction from low-resolution inputs remains a significant challenge. To address this, we introduce a learnable downsampler, which allows the model to better capture diverse degradation patterns and enhance SR performance.

In generative tasks, standard diffusion models have demonstrated impressive capabilities in producing high-quality images. However, their slow, iterative sampling process limits their applicability to real-time scenarios. To address this limitation, various acceleration techniques have been proposed. Methods like DPM-solver~\cite{Lu2022} and DDIM~\cite{DDIM} reduce the number of sampling steps, but they often sacrifice image quality, leading to blurry results. As such, subsequent research has focused on improving both efficiency and quality in generative tasks.

One promising direction for enhancing efficiency is distillation-based acceleration, where a student model learns to replicate the teacher model’s output in fewer steps. Techniques like Progressive Distillation~\cite{Salimans2022} and Guided Distillation~\cite{Meng2023} significantly reduce computational costs. However, these methods still require either prolonged training or multiple sampling steps, which limits their scalability for real-time SR applications.

Another emerging approach combines diffusion models with Generative Adversarial Networks (GANs) to address both efficiency and output quality. Methods such as Denoising Diffusion GANs~\cite{Xiao2022}, UFOGen~\cite{Xu2024}, and ADD~\cite{ADD} demonstrate the potential of combining diffusion models with discriminators to enhance the generative process. These hybrid approaches offer promising results by balancing the strengths of both diffusion models and GANs.

\begin{figure}[]
    \centering
    \includegraphics[width=1\linewidth]{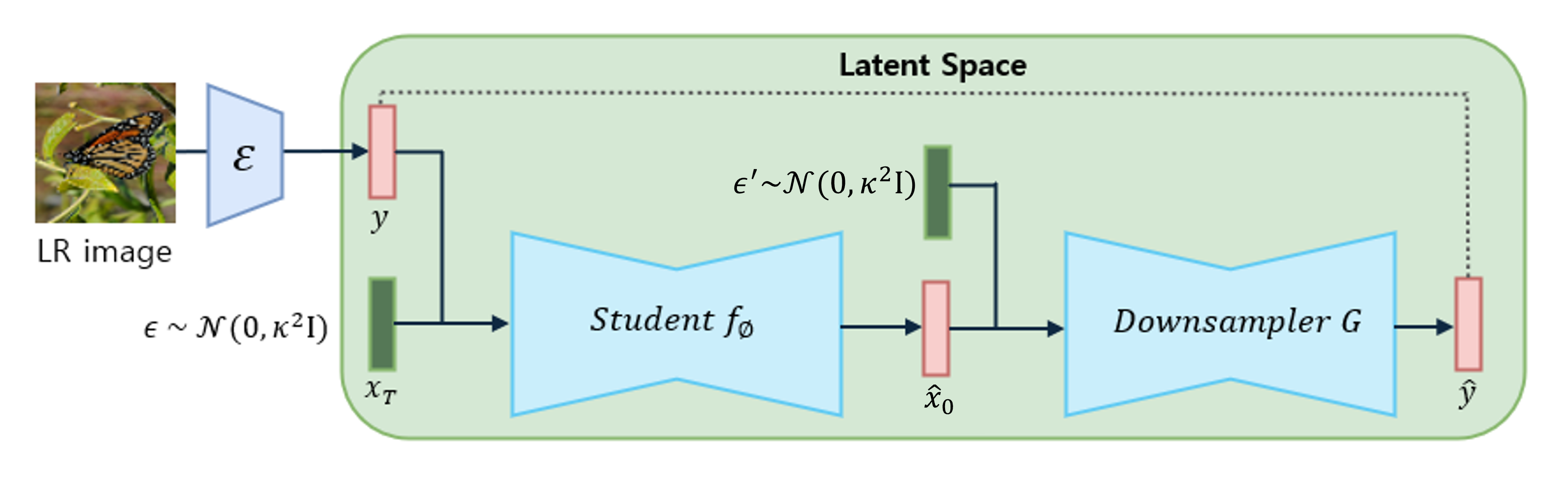}
    \caption{Both the student network (low-to-high) and the downsampler (high-to-low) are diffusion-based architectures. In the latent space, the output of the student network is conditioned and fed into the downsampler. The output of the downsampler $\hat{y}$ is then compared with the original low-resolution $y$ during training.}
    \label{fig:enter-label}
\end{figure}

Building upon these advancements, we propose a novel single-step diffusion model for SR, distilled from a pre-trained teacher model and augmented with two discriminators. Our key innovation is the introduction of a learnable downsampler, which incorporates realistic degradation patterns into the training process, thus improving SR performance. The two discriminators play distinct yet complementary roles. \textit{(i) HR discriminator}: Uses ground-truth high-resolution images to refine and improve the generated image quality. \textit{(ii) LR discriminator}: Employs a flexible diffusion-based downsampler to model the degradation process, comparing the generated LR images with the corresponding low-resolution inputs.

Our contributions are summarized as follows: \begin{itemize} \item A novel method combining diffusion networks with two discriminators for SR tasks. \item Both the upsampler and downsampler are diffusion networks that are learned via their interaction in the latent space. \item Achieving state-of-the-art performance in a single-step SR process, without increasing the complexity of the student network. \item Demonstrating superior performance on real-world SR tasks, showcasing the effectiveness of our method. \end{itemize}

\section{Related Work}
\label{sec:relatedwork}

\subsection{Diffusion model}
Diffusion models~\cite{DDPM} are generative models that gradually transform noise into data through a denoising process. It starts with noise $x_T$ and gradually produces less noisy samples $x_{T-1}, x_{T-2}, \ldots, x_0$. Recently, diffusion models~\cite{Rombach2022} have shown state-of-the-art performance in various domains, including image generation and image super-resolution (SR).


\subsection{Image Super-Resolution}
Super-Resolution(SR) restores a high-resolution(HR) image from a low-resolution(LR) input. Early works in SR using Deep Neural Networks (DNNs) include SRCNN~\cite{Dong2015}, which was one of the first deep learning approaches to show significant improvements over traditional interpolation-based techniques. These models directly map LR images to HR outputs, often with a focus on pixel-wise reconstruction quality.


\paragraph{GAN-based models}
Generative Adversarial Networks (GANs)\cite{Goodfellow2014} have been widely applied to Image Super-Resolution (ISR). SRGAN\cite{Ledig2017} was one of the first to demonstrate the effectiveness of GANs for SR, using adversarial loss. Over time, more sophisticated approaches such as BSRGAN~\cite{degradation2} and Real-ESRGAN~\cite{Wang2021} have emerged, specifically designed to handle real-world degradation patterns and improve perceptual quality.

\paragraph{Diffusion-based models}
Diffusion models have also been increasingly applied to Super-Resolution~\cite{kim2024arbitrary, Choi2021}. These methods typically either directly incorporate the LR image into the denoising network's input~\cite{Rombach2022} or use pre-trained models~\cite{Choi2021} to generate HR images. The key advantage of diffusion-based SR methods is their ability to model high-level image structures and textures in addition to fine details. However, despite their strong performance, the efficiency of diffusion models is often limited by the number of inference steps required.

\paragraph{Downsampler}
Learnable downsamplers \cite{degradation1, aakerberg2024pda, wang2023lldiffusion} have been employed to generate LR-HR pairs for training with GAN or enhance degradation features for super-resolution. In some works, \cite{menon2020pulse} penalizes super-resolved images that deviate from LR inputs, while \cite{featup} proposes a learnable downsampler that directly influences the upsampler.


\section{Preliminary}
\subsection{Deterministic sampling}
SinSR~\cite{SinSR} introduces a non-Markovian reverse process, enabling deterministic sampling and distillation in a single step. Inspired by DDIM, the reverse process conditioned on a given LR image $y$ is defined as
\begin{equation} x_{t-1} = k_t \hat{x}_0 + m_t x_t + j_t y, \label{eq:5} \end{equation}
where $\hat{x}_0 = f_{\theta}(x_t, y, t)$ represents the high-resolution image predicted by a pre-trained teacher model. The coefficients $k_t$, $m_t$, and $j_t$ are derived based on a shifting sequence $\eta_t$.

\section{Methodology}
\label{sec:3}

\begin{figure}[]
  \centering
  \includegraphics[width=0.88\columnwidth]{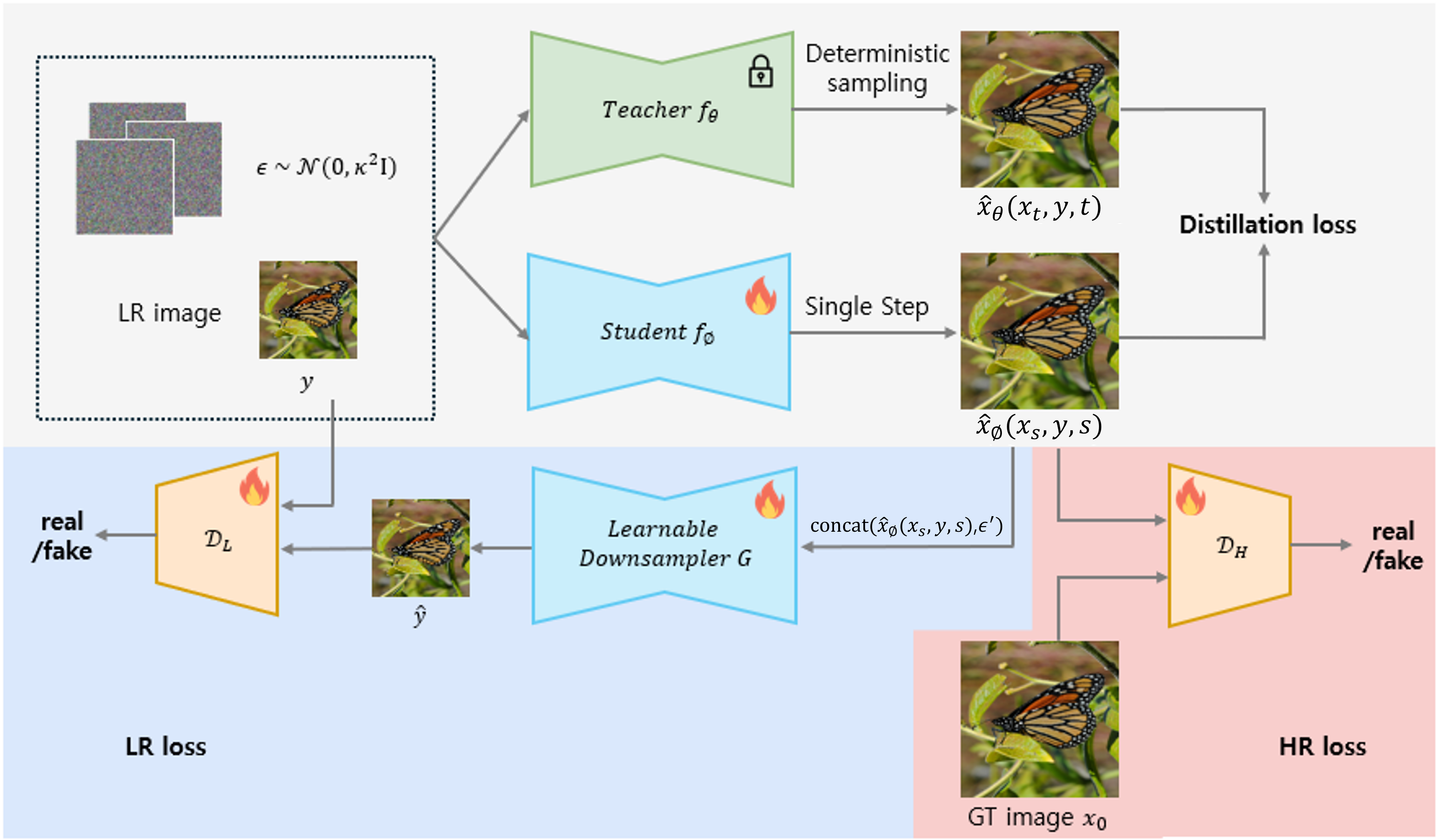}
  \caption{The overall framework of our model. The student network $f_{\phi}$ is trained to learn a deterministic mapping from $x_T$ to $\hat{x}_0$ in just one step, guided by a pre-trained teacher network $f_\theta$. The student's output $\hat{x}_\phi(x_s,y,s)$ then goes to the High-Resolution Discriminator $\mathcal{D}_H$. Simultaneously, $\hat{x}_\phi(x_s,y,s)$ is jointly learnt with a Learnable Downsampler $G$ and Low-Resolution Discriminator $\mathcal{D}_L$ in end-to-end fashion.} 
  \label{fig:architecture_TDDSR}
\end{figure}

Our training process, illustrated in Fig.~\ref{fig:architecture_TDDSR}, involves three key components and five networks: the teacher model with frozen weights $\theta$, the student model with weight $\phi$, a learnable downsampler \textit{G}, a high-resolution(HR) discriminator \textit{H}, and a low-resolution(LR) discriminator \textit{L}. First, knowledge from the teacher model is distilled into the student model, starting with the initial state $x_T=y+\kappa\sqrt{\eta_T}\epsilon$, where $\epsilon \sim \mathcal{N}(\mathbf{0},\mathbf{I})$. Second, the student's output, $\hat{x}_\phi(x_s,y,s)$, is evaluated by the HR discriminator, which distinguishes it from the ground-truth high-resolution image $x_0$. Lastly, the student’s output is passed through the learnable downsampler to generate an estimated low-resolution image $\hat{y}$, which is then assessed by the LR discriminator against the original low-resolution input $y$. In our experiments, for the final timestep $T$, the teacher model's timestep $t$ progresses up to $T_{\text{teacher}}=15$, while the student model's timestep $s$ operates in a single step, reaching $T_{\text{student}}=1$.
Inspired by DiscoGAN~\cite{DiscoGAN} and CycleGAN~\cite{CycleGAN}, we leverage cycle consistency to enforce bidirectional learning between LR and HR domains. While conventional SR methods primarily focus on LR-to-HR mapping, our approach integrates a learnable downsampler to model the HR-to-LR degradation. During training, the diffusion-based framework generates diverse high- and low-resolution samples at each iteration. In particular, the diffusion-based downsampler applies varying degradation patterns, enriching the learning process and enhancing the student model's generalization.

\subsection{Distillation}
\label{sec:3.1}
We adapt the concepts of SinSR\cite{SinSR}. Using the teacher network $f_\theta$ and the student network $f_\phi$, the teacher model iterates over timesteps $t \in \{ 1,...,T_{\text{teacher}} \}$, as shown in Eq.~\ref{eq:5}, yielding the output $\hat{x}_\theta(x_t,y,t)$. Similarly, the student model generates $\hat{x}_\phi(x_s,y,s)$ with $s \in \{1,..., T_{\text{student}}\}$. The distillation loss is formulated as
\begin{equation}
    \mathcal{L}_{\text{Distill}} = L_{\text{MSE}} \left( \hat{x}_\phi(x_s,y,s), \hat{x}_\theta(x_t,y,t) \right),
    \label{eq:6}
\end{equation}
where $T_{\text{teacher}}=15$ and $T_{\text{student}}=1$, meaning the student network $f_{\phi}$ learns a deterministic mapping between the initial state $x_T=y+\kappa\sqrt{\eta_T}\epsilon$ and the teacher's estimated output $\hat{x}_\theta(x_t,y,t)$, distilled to predict the high-resolution (HR) image in just one step.

\subsection{High-Resolution Image Discriminator}
\label{sec:3.2}

In distillation, the student network learns super-resolution in a single step by minimizing the distillation loss $\mathcal{L}_{\text{Distill}}$, comparing its output with the teacher's. However, the teacher's output alone cannot fully capture real-world degradation complexities, limiting performance. To overcome this, we use Ground-truth (GT) images during training, which helps reduce artifacts and improve generalization. Inspired by ADD~\cite{ADD}, we integrate a discriminator to directly compare the student's output with the GT images.

For the discriminator, we adopted the design proposed in StyleGAN-T~\cite{Stylegan-t}, utilizing and training a feature network. During the adversarial training, the high-resolution discriminator $\mathcal{D}_{H}$ is trained to minimize 
\begin{equation}
\begin{aligned}
\mathcal{L}_{\mathcal{D}_H} &= \mathbb{E}_{x_0} \left[ \log(1 - \mathcal{D}_H (x_0)) \right] \\
              &\quad + \mathbb{E}_{\hat{x}_\phi(x_s,y,s)} \left[ \log(\mathcal{D}_H (\hat{x}_\phi(x_s,y,s))) \right],
\end{aligned}
\label{eq:7}
\end{equation}
where $x_0$ is the ground-truth HR image and $\hat{x}_\phi(x_s,y,s)$ is the student's predicted high-resolution image. The student network (e.g., the generator) is optimized using the loss function as follows
\begin{equation}
\mathcal{L}^G_H = \mathbb{E}_{\hat{x}_\phi(x_s,y,s)} \left[ 1 - \mathcal{D}_H (\hat{x}_\phi(x_s,y,s)) \right].
\label{eq:8}
\end{equation}

\subsection{Learnable Downsampler and Low-Resolution Image Discriminator}
\label{sec:3.3}

\paragraph{Learnable Downsampler}
Our method stands out from prior works through its backbone architecture, which integrates a diffusion-based approach with a flexible learnable downsampler. This downsampler captures diverse degradation patterns and generates varying low-resolution (LR) images during training. Sharing the same structure as the student network, it adapts to different degradation scenarios, providing richer information that enhances the student network’s performance. Both networks are diffusion-based, with the student network’s output concatenated with noise in the latent space and fed into the downsampler, as shown in Fig.~\ref{fig:enter-label}. The Student receives a low-resolution image $y$ conditioned on Gaussian noise $\epsilon \sim \mathcal{N}(0, \kappa^2 \mathbf{I})$, forming a noisy input $x_T = y + \epsilon$. It then predicts a high-resolution latent representation $\hat{x}_0 = \hat{x}_\phi(x_s,y,s)$, which is concatenated with additional noise $\epsilon' \sim \mathcal{N}(0, \kappa^2 \mathbf{I})$. This concatenated input is passed to the downsampler $G$, which reconstructs the low-resolution image $\hat{y}=G(\text{concat}(\hat{x}_\phi(x_s,y,s), \epsilon'))$. The concatenation mechanism facilitates communication between the two diffusion networks.

\paragraph{Low-Resolution Image Discriminator}
The Low-Resolution Image Discriminator $\mathcal{D}_L$ utilizes the same architecture and loss function as in Sec~\ref{sec:3.2}for consistency and simplified implementation. 
The low-resolution discriminator $\mathcal{D}_L$ is trained to minimize
\begin{equation}
\begin{aligned}
\mathcal{L}_{\mathcal{D}_L} &= \mathbb{E}_{y} \left[ \log(1 - \mathcal{D}_L (y)) \right] \\
              &\quad + \mathbb{E}_{\hat{x}_\phi(x_s,y,s)} \left[ \log(\mathcal{D}_L (G(\text{concat}(\hat{x}_\phi(x_s,y,s), \epsilon'))) \right],
\end{aligned}
\label{eq:9}
\end{equation}
where $y$ is the ground-truth low-resolution image, and $\hat{y}=G(\text{concat}(\hat{x}_\phi(x_s,y,s), \epsilon'))$ is the LR image predicted by the learnable downsampler $G$. The downsampler's adversarial objective amounts to
\begin{equation}
\begin{aligned}
\mathcal{L}^G_L &= \mathbb{E}_{\hat{x}_\phi(x_s,y,s)} \left[ 1 - \mathcal{D}_L (G(\text{concat}(\hat{x}_\phi(x_s,y,s), \epsilon'))) \right].
\end{aligned}
\label{eq:10}
\end{equation}


\begin{table*}[!htbp]
\centering
\vspace{4pt}
\resizebox{0.98\textwidth}{!}{
\begin{tabular}{c|ccccccccc}
\hline
\multirow{3}{*}{Methods}    & \multicolumn{9}{c}{Datasets}                                                                                                                                                                                           \\ \cline{2-10} 
                            & \multicolumn{2}{c|}{\textit{RealSR}}                      & \multicolumn{2}{c|}{\textit{RealSet65}}                   & \multicolumn{5}{c}{\textit{DIV2K-Val}}                                                         \\ \cline{2-10} 
                            & CLIPIQA(↑)       & \multicolumn{1}{c|}{MUSIQ(↑)}          & CLIPIQA(↑)       & \multicolumn{1}{c|}{MUSIQ(↑)}          & PSNR(↑)           & SSIM(↑)          & LPIPS(↓)         & CLIPIQA(↑)       & MUSIQ(↑)          \\ \hline
RankSRGAN                   & 0.582            & \multicolumn{1}{c|}{\underline {62.098}} & 0.560            & \multicolumn{1}{c|}{51.813}            & 26.510            & 0.753            & \textbf{0.122}   & 0.640            & 64.686            \\
Real-ESRGAN                 & 0.490            & \multicolumn{1}{c|}{59.692}            & 0.599            & \multicolumn{1}{c|}{\underline {63.220}} & 26.651            & \underline {0.758} & 0.228            & 0.565            & 64.655            \\
LDM(15 sampling steps)      & 0.384            & \multicolumn{1}{c|}{49.317}            & 0.427            & \multicolumn{1}{c|}{47.488}            & 25.587            & 0.685            & 0.234            & 0.668            & 65.047            \\
ResShift(15 sampling steps) & 0.601            & \multicolumn{1}{c|}{58.648}            & 0.649            & \multicolumn{1}{c|}{60.772}            & \textbf{27.075}   & \textbf{0.763}   & 0.201            & 0.673            & 65.570            \\
SinSR(single step)          & \underline {0.686} & \multicolumn{1}{c|}{60.750}            & \underline {0.715} & \multicolumn{1}{c|}{62.258}            & 26.622            & 0.732            & 0.207            & \textbf{0.715}   & \underline {65.764} \\ \hline
Ours(single step)           & \textbf{0.724}   & \multicolumn{1}{c|}{\textbf{63.263}}   & \textbf{0.743}   & \multicolumn{1}{c|}{\textbf{64.063}}   & \underline {26.880} & 0.747            & \underline {0.198} & \underline {0.686} & \textbf{65.865}   \\ \hline
\end{tabular}
}
\caption{Quantitative results on the real-world datasets(RealSR and RealSet65) and DIV2K validation dataset. The best and second best results are highlighted in \textbf{bold} and \underline{underlined}.}
\label{tab:real_div_results}
\end{table*}
\subsection{Losses for Two Diffusion Networks}
\label{sec:3.4}
Both the student network (low-to-high) and the downsampler network (high-to-low) employ diffusion models, with their respective total losses defined as follows. With $\mathcal{L}_{denoise}=\mathbb{E}[||\epsilon - \epsilon_\theta(x_t) ||^2_2]=\frac{\bar{\alpha}_t}{1-\bar{\alpha}_t}||x_0-\hat{x}_0 ||^2_2$, $\bar{\alpha}_t := \sum_{k=1}^t (1-\beta_t)$, $\{ \beta_t \}_{t=0}^T$ is a variance schedule, and $\theta$ representing the model parameters, the total loss of the student network is defined as follows
\begin{equation}
\begin{aligned}
\mathcal{L}_{student} &= \mathcal{L}_{denoise} + \frac{\bar{\alpha}_t}{1-\bar{\alpha}_t}(\mathcal{L}_{distill} + \mathcal{L}^G_H + \mathcal{L}^G_L ).
\end{aligned}
\label{eq:11}
\end{equation}
The total loss of the downsampler is defined as follows
\begin{equation}
\begin{aligned}
\mathcal{L}_{downsampler} &= \mathcal{L}_{denoise} + \frac{\bar{\alpha}_t}{1-\bar{\alpha}_t} \mathcal{L}^G_L .
\end{aligned}
\label{eq:12}
\end{equation}


\section{Experiments}

We trained our model using the Adam optimizer with a learning rate of 5e-5 for 20k iterations. All experiments were conducted on a single NVIDIA A100 GPU with 40GB of memory. Our evaluation covers both Real Image Super-Resolution (Real-ISR), which includes a diverse range of real-world images, and Face Super-Resolution (Face-SR), which focuses on domain-specific face datasets. To ensure task-appropriate evaluations, we used different datasets and metrics for each category. Our study specifically targets the challenging $\times$4 super-resolution task. Additionally, we utilized ResShift~\cite{ResShift} as the teacher model, which was pre-trained on ImageNet.


\subsection{Real-world Super-Resolution}

\paragraph{Datasets}
For training, we use the DIV2K dataset~\cite{agustsson2017ntire}, degraded with the RealESRGAN~\cite{Wang2021} pipeline, following the approach in ResShift~\cite{ResShift}. To assess generalization on unseen real-world data, we evaluate our model on RealSR~\cite{cai2019toward} and RealSet65~\cite{ResShift}. RealSR consists of 100 real images captured with Canon and Nikon cameras in diverse settings, while RealSet65 includes 65 images sourced from widely known datasets and online sources.

\paragraph{Metrics and Compared Methods}
We use PSNR, SSIM, and LPIPS~\cite{zhang2018unreasonable} as reference metrics to assess fidelity and perceptual quality. For non-reference evaluation, we employ CLIPIQA~\cite{wang2023exploring} and MUSIQ~\cite{ke2021musiq}, two recently introduced metrics, to measure the realism of generated images, particularly for real-world datasets. To demonstrate the effectiveness of our model, we compare it against several state-of-the-art SR methods, including SinSR~\cite{SinSR}, ResShift~\cite{ResShift}, LDM(Latent Diffusion Model)~\cite{Rombach2022}, Real-ESRGAN~\cite{Wang2021}, and RankSRGAN~\cite{zhang2019ranksrgan}, to demonstrate its effectiveness.

\begin{figure}[]
    \centering
    \resizebox{\columnwidth}{!}{
    \begin{subfigure}{0.2\columnwidth}
        \includegraphics[width=\linewidth]{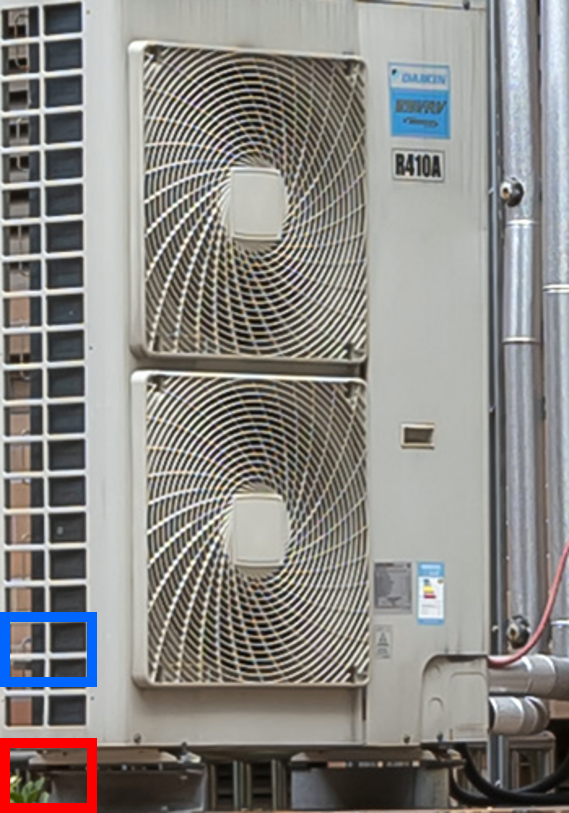}
    \end{subfigure}
    \hfill
    \begin{subfigure}{0.197\columnwidth}
        \includegraphics[width=\linewidth]{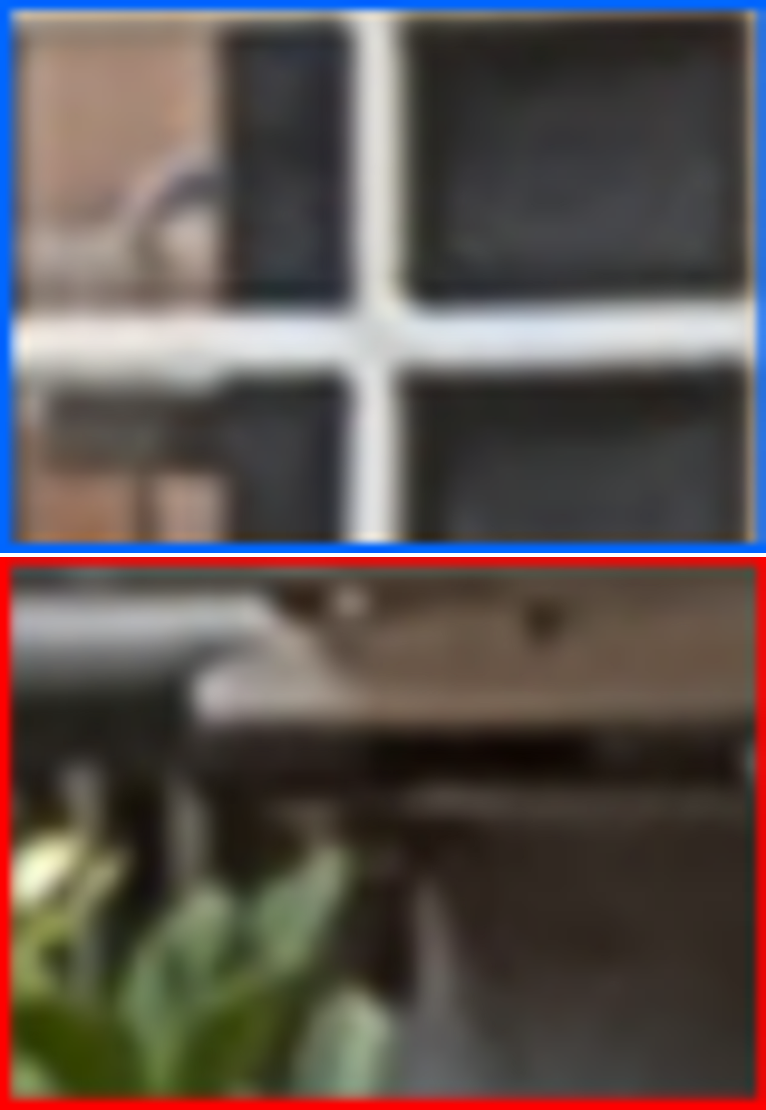}
    \end{subfigure}
    \hfill
    \begin{subfigure}{0.19\columnwidth}
        \includegraphics[width=\linewidth]{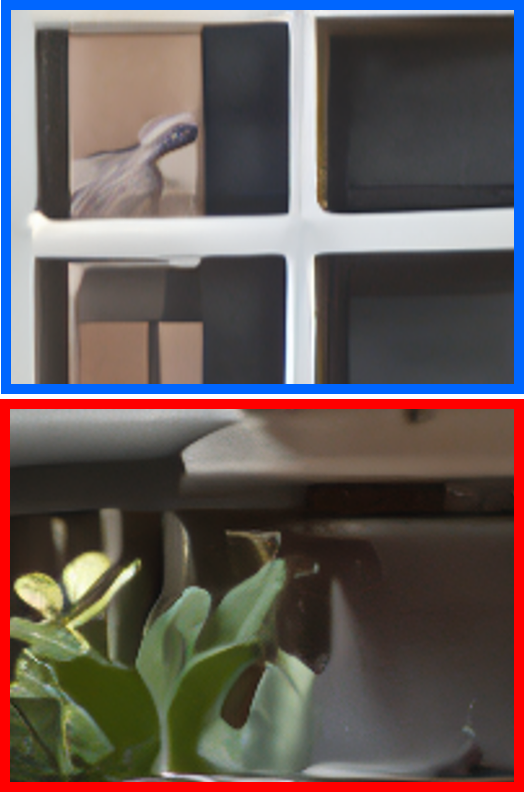}
    \end{subfigure}
    \hfill
    \begin{subfigure}{0.19\columnwidth}
        \includegraphics[width=\linewidth]{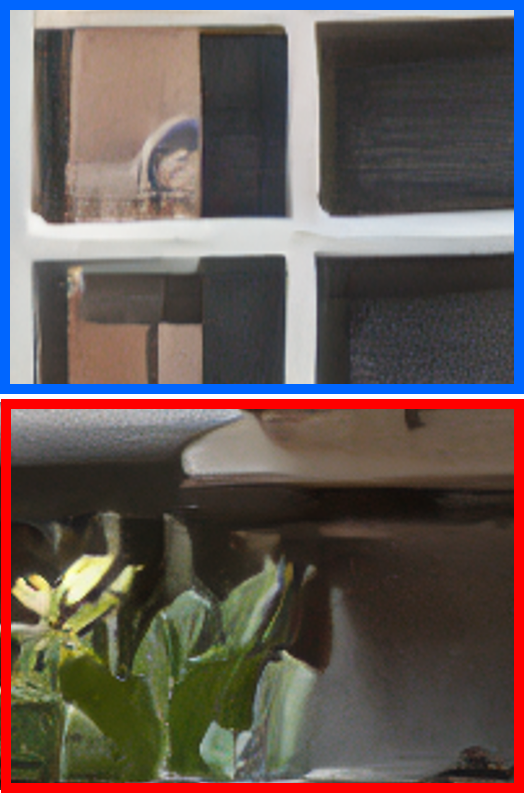}
    \end{subfigure}
    \hfill
    \begin{subfigure}{0.193\columnwidth}
        \includegraphics[width=\linewidth]{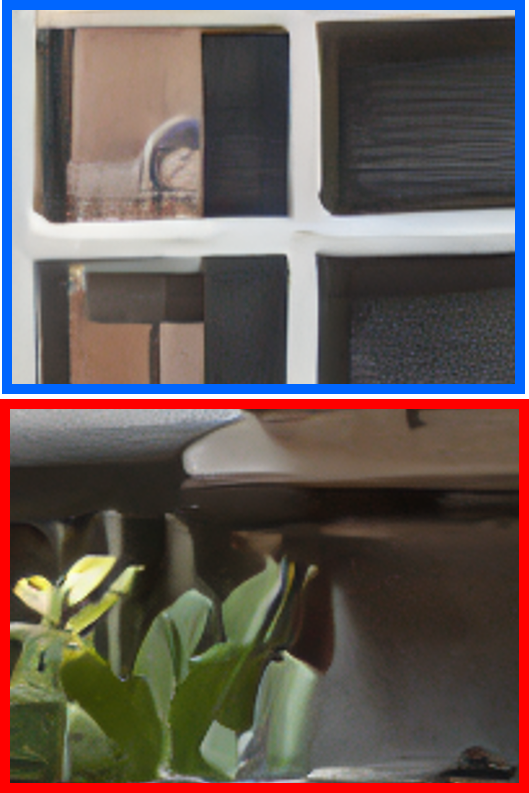}
    \end{subfigure} 
    }

    \vspace{0.05cm}
    
    \resizebox{\columnwidth}{!}{
    \begin{subfigure}{0.24\columnwidth}
        \includegraphics[width=\linewidth]{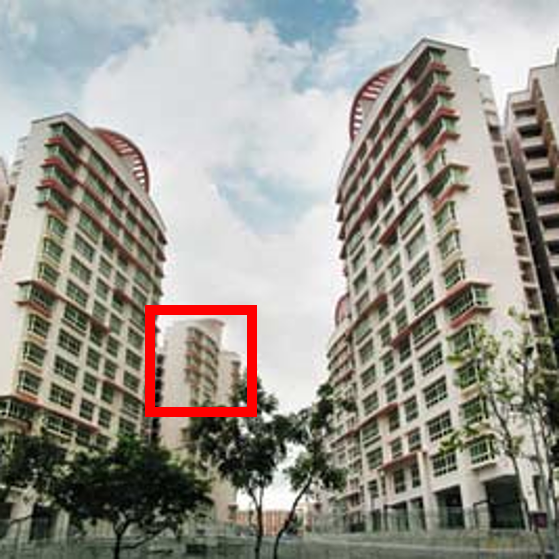}
    \end{subfigure}
    \hfill
    \begin{subfigure}{0.24\columnwidth}
        \includegraphics[width=\linewidth]{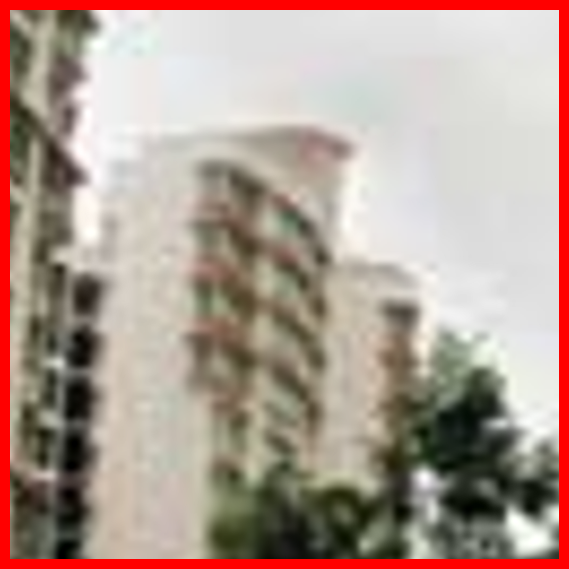}
    \end{subfigure}
    \hfill
    \begin{subfigure}{0.24\columnwidth}
        \includegraphics[width=\linewidth]{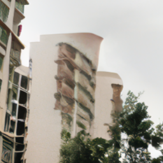}
    \end{subfigure}
    \hfill
    \begin{subfigure}{0.24\columnwidth}
        \includegraphics[width=\linewidth]{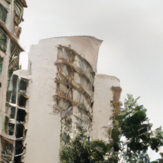}
    \end{subfigure}
    \hfill
    \begin{subfigure}{0.24\columnwidth}
        \includegraphics[width=\linewidth]{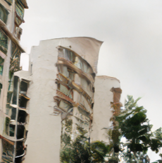}
    \end{subfigure}
    }

    \vspace{0.05cm}
    
    \resizebox{\columnwidth}{!}{
    \begin{subfigure}{0.24\columnwidth}
        \includegraphics[width=\linewidth]{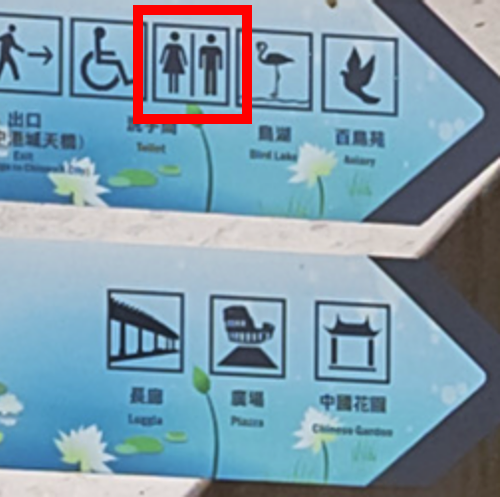}
        \caption{Full LR}
    \end{subfigure}
    \hfill
    \begin{subfigure}{0.24\columnwidth}
        \includegraphics[width=\linewidth]{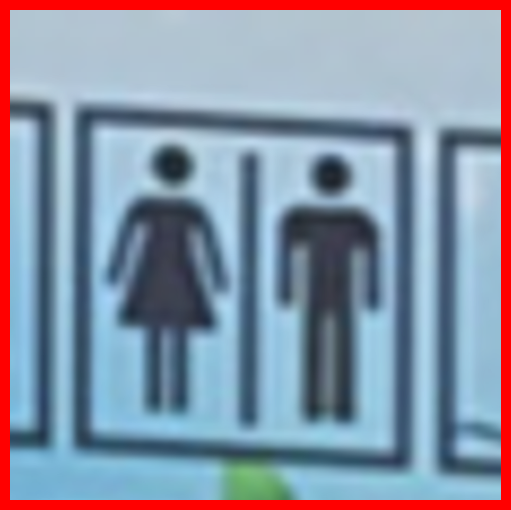}
        \caption{Zoomed LR}
    \end{subfigure}
    \hfill
    \begin{subfigure}{0.24\columnwidth}
        \includegraphics[width=\linewidth]{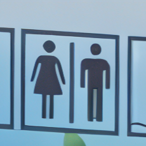}
        \caption{ResShift-15}
    \end{subfigure}
    \hfill
    \begin{subfigure}{0.24\columnwidth}
        \includegraphics[width=\linewidth]{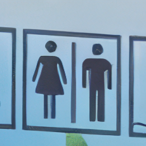}
        \caption{SinSR-1}
    \end{subfigure}
    \hfill
    \begin{subfigure}{0.24\columnwidth}
        \includegraphics[width=\linewidth]{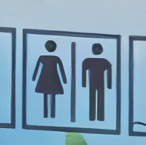}
        \caption{Ours-1}
    \end{subfigure}
    }
    \caption{Visual comparisons on real-world datasets[RealSR, RealSet65]. Zoom in for more details.}
    \label{fig:comparison}
\end{figure}

\begin{figure}[]
    \centering
    \resizebox{\columnwidth}{!}{
    \begin{subfigure}{0.24\columnwidth}
        \includegraphics[width=\linewidth]{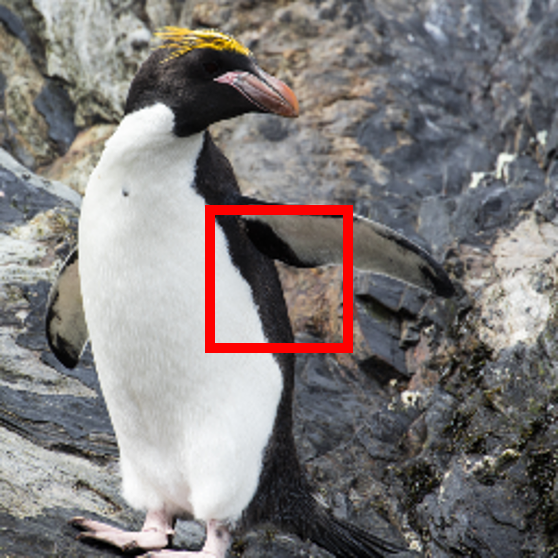}
    \end{subfigure}
    \hfill
    \begin{subfigure}{0.24\columnwidth}
        \includegraphics[width=\linewidth]{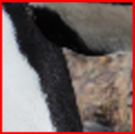}
    \end{subfigure}
    \hfill
    \begin{subfigure}{0.24\columnwidth}
        \includegraphics[width=\linewidth]{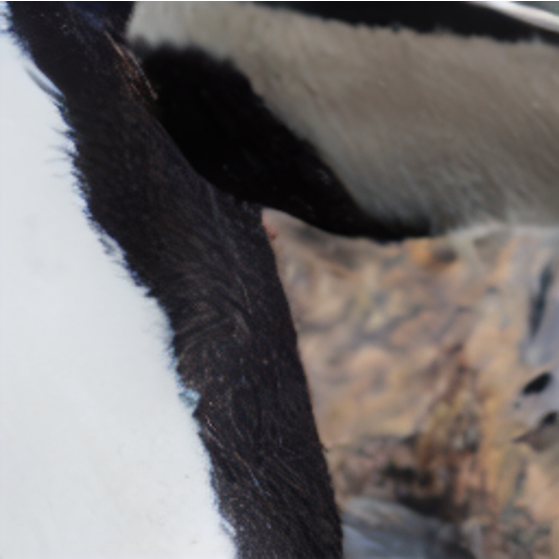}
    \end{subfigure}
    \hfill
    \begin{subfigure}{0.24\columnwidth}
        \includegraphics[width=\linewidth]{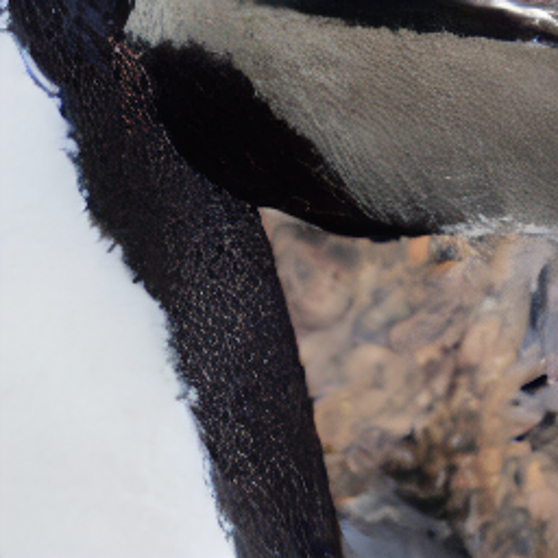}
    \end{subfigure}
    \hfill
    \begin{subfigure}{0.24\columnwidth}
        \includegraphics[width=\linewidth]{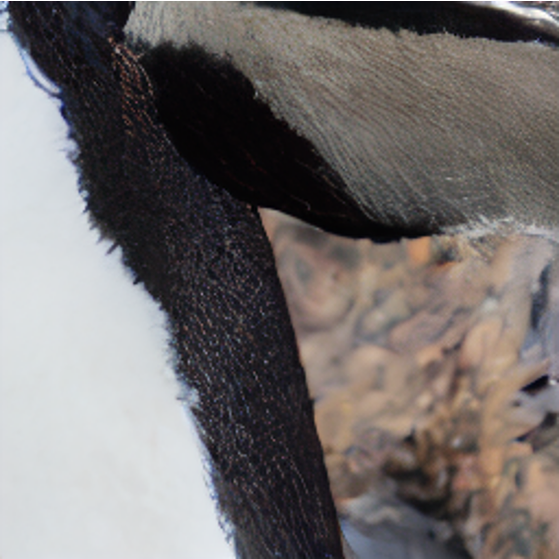}
    \end{subfigure}
    }

    \vspace{0.05cm}
    
    \resizebox{\columnwidth}{!}{
    \begin{subfigure}{0.24\columnwidth}
        \includegraphics[width=\linewidth]{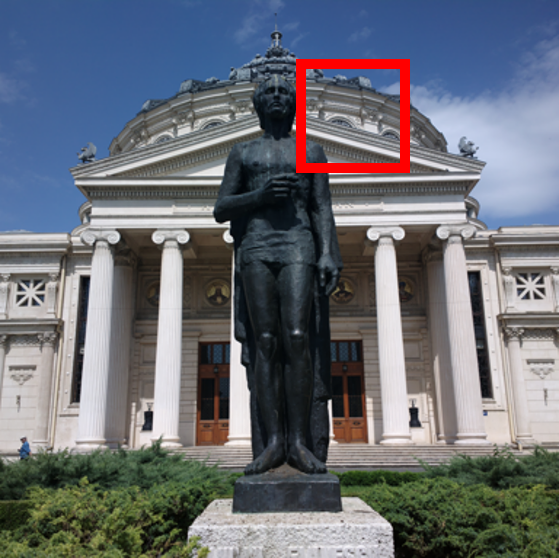}
    \end{subfigure}
    \hfill
    \begin{subfigure}{0.24\columnwidth}
        \includegraphics[width=\linewidth]{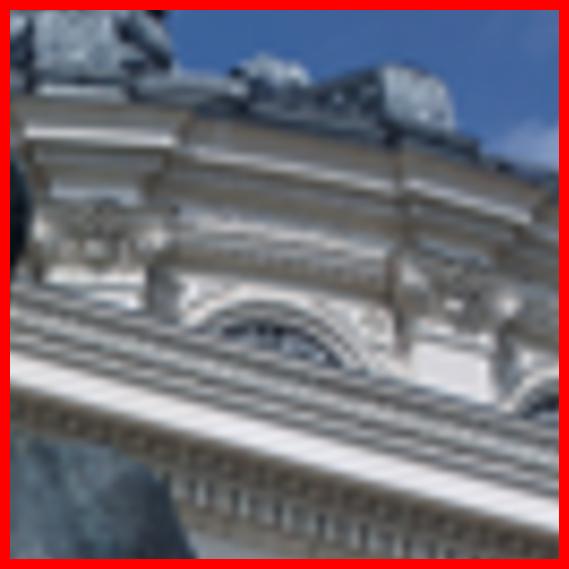}
    \end{subfigure}
    \hfill
    \begin{subfigure}{0.24\columnwidth}
        \includegraphics[width=\linewidth]{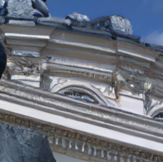}
    \end{subfigure}
    \hfill
    \begin{subfigure}{0.24\columnwidth}
        \includegraphics[width=\linewidth]{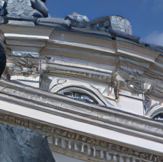}
    \end{subfigure}
    \hfill
    \begin{subfigure}{0.24\columnwidth}
        \includegraphics[width=\linewidth]{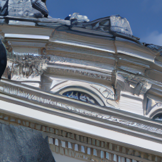}
    \end{subfigure}
    }

    \vspace{0.05cm}
    
    \resizebox{\columnwidth}{!}{
    \begin{subfigure}{0.24\columnwidth}
        \includegraphics[width=\linewidth]{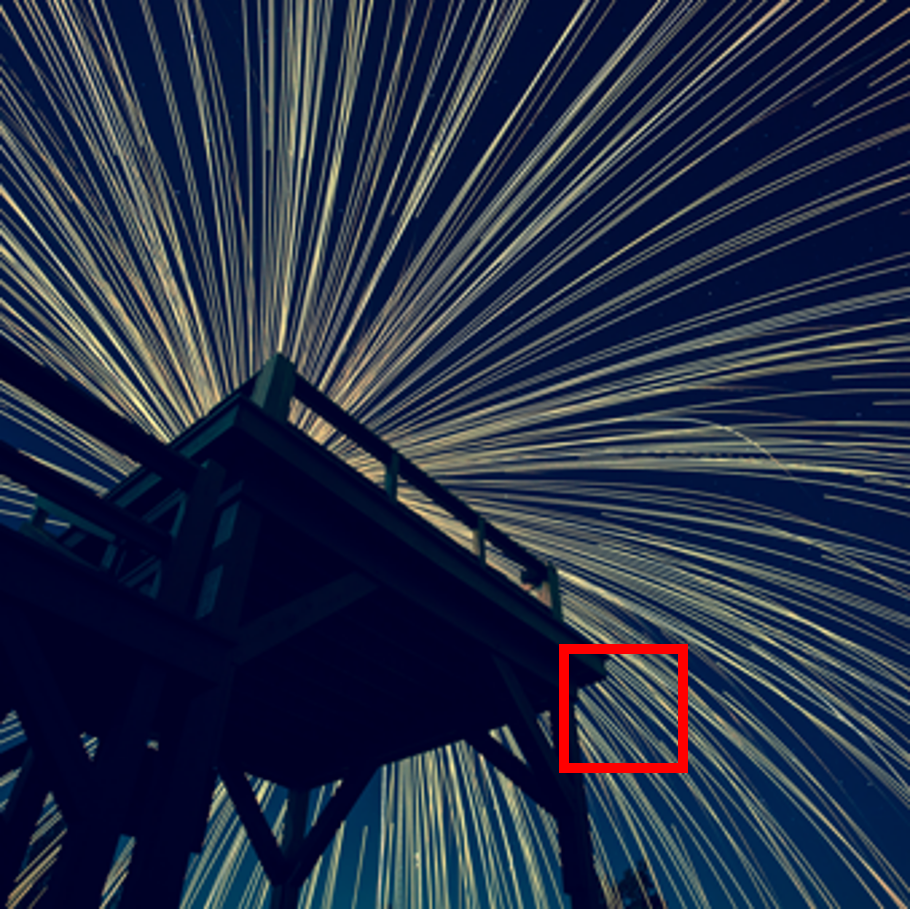}
        \caption{Full LR}
    \end{subfigure}
    \hfill
    \begin{subfigure}{0.24\columnwidth}
        \includegraphics[width=\linewidth]{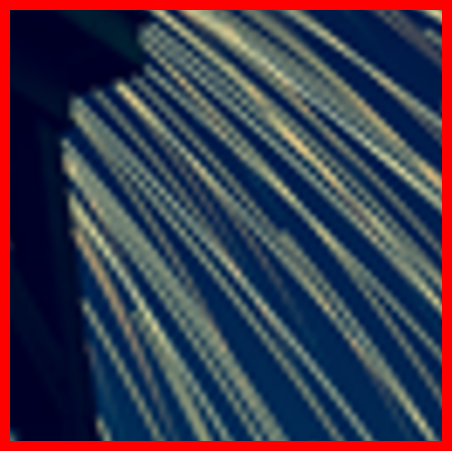}
        \caption{Zoomed LR}
    \end{subfigure}
    \hfill
    \begin{subfigure}{0.24\columnwidth}
        \includegraphics[width=\linewidth]{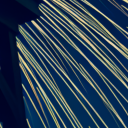}
        \caption{ResShift-15}
    \end{subfigure}
    \hfill
    \begin{subfigure}{0.24\columnwidth}
        \includegraphics[width=\linewidth]{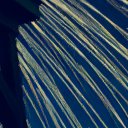}
        \caption{SinSR-1}
    \end{subfigure}
    \hfill
    \begin{subfigure}{0.24\columnwidth}
        \includegraphics[width=\linewidth]{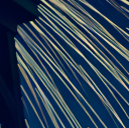}
        \caption{Ours-1}
    \end{subfigure}
    }
    \caption{Visual comparisons on DIV2K-Val dataset. Zoom in for more details.}
    \label{fig:comparison2}
\end{figure}

\paragraph{Results}
The results for both real-world datasets (RealSR, RealSet65) and the DIV2K validation dataset, with 256$\times$256 outputs, are presented in Table~\ref{tab:real_div_results}. For real-world datasets, our proposed approach outperforms SinSR and the teacher model, highlighting its effectiveness in real-world scenarios. On the DIV2K validation dataset, despite a drop in PSNR and SSIM due to reducing inference steps from 15 to 1, our model achieves superior perceptual quality with higher LPIPS, CLIPIQA, and MUSIQ scores, effectively balancing efficiency and visual fidelity. Additionally, Ours(single-step) outperforms SinSR(single-step) across all metrics except CLIPIQA. 
Visual comparisons across datasets are shown in Fig.\ref{fig:comparison} and Fig.\ref{fig:comparison2}, illustrating the qualitative differences among the teacher model ResShift, the single-step model SinSR, and our proposed single-step model. The '-N' following each method's name denotes the number of inference steps. Our model produces results that are less blurry and exhibit fewer artifacts.

\subsection{Face-specific Super-Resolution}
\paragraph{Datasets}
We utilize the Flick-Faces-HQ (FFHQ) dataset~\cite{karras2019style}, which contains a diverse collection of 70,000 high-quality human face images. We partition this dataset into non-overlapping subsets: 80\% for training, 15\% for testing, and 5\% for validation. 

\paragraph{Metrics and Compared Methods}
To quantitatively assess the performance of our face SR methods, we use evaluation metrics including PSNR, SSIM, Multi-Scale SSIM (MS-SSIM)\cite{wang2003multiscale}, and Fréchet Inception Distance (FID). We compare our approach against several state-of-the-art methods, including SRGAN~\cite{Ledig2017}, ESRGAN~\cite{Wang2018}, EnhanceNet~\cite{Sajjadi2017}, SRFBN~\cite{li2019feedback}, CAGFace~\cite{kalarot2020component}, and SinSR~\cite{SinSR}. We compare our results with those reported in \cite{kalarot2020component} for a direct benchmark.

\begin{figure}[]
    \centering
    \begin{subfigure}{0.94\columnwidth}
        \centering
        \begin{subfigure}{0.24\columnwidth}
            \includegraphics[width=\linewidth]{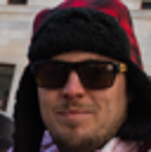}
        \end{subfigure}
        \begin{subfigure}{0.24\columnwidth}
            \includegraphics[width=\linewidth]{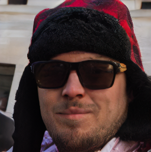}
        \end{subfigure}
        \begin{subfigure}{0.24\columnwidth}
            \includegraphics[width=\linewidth]{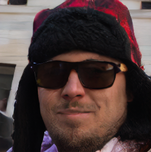}
        \end{subfigure}   
        \begin{subfigure}{0.24\columnwidth}
            \includegraphics[width=\linewidth]{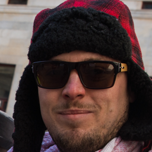}
        \end{subfigure} 
    \end{subfigure}

    \vspace{0.1cm}
    
    \begin{subfigure}{0.94\columnwidth}
        \centering
        \begin{subfigure}{0.24\columnwidth}
            \includegraphics[width=\linewidth]{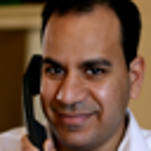}
        \end{subfigure}
        \begin{subfigure}{0.24\columnwidth}
            \includegraphics[width=\linewidth]{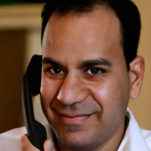}
        \end{subfigure}
        \begin{subfigure}{0.24\columnwidth}
            \includegraphics[width=\linewidth]{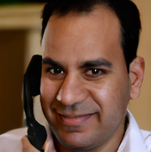}
        \end{subfigure}   
        \begin{subfigure}{0.24\columnwidth}
            \includegraphics[width=\linewidth]{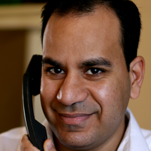}
        \end{subfigure}  
    \end{subfigure}

    \vspace{0.1cm}
    
    \begin{subfigure}{0.94\columnwidth}
        \centering
        \begin{subfigure}{0.24\columnwidth}
            \includegraphics[width=\linewidth]{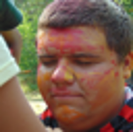}
            \caption{LR input}
        \end{subfigure}
        \begin{subfigure}{0.24\columnwidth}
            \includegraphics[width=\linewidth]{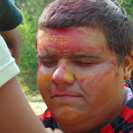}
            \caption{Ours}
        \end{subfigure}
        \begin{subfigure}{0.24\columnwidth}
            \includegraphics[width=\linewidth]{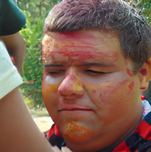}
            \caption{SinSR}
        \end{subfigure}  
        \begin{subfigure}{0.24\columnwidth}
            \includegraphics[width=\linewidth]{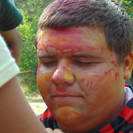}
            \caption{GT}
        \end{subfigure} 
    \end{subfigure}
    \caption{Comparisons on FFHQ dataset where 64$\times$64 inputs are upscaled to 256$\times$256 high-resolution outputs(4$\times$). Our results are compared with SinSR, a single-step diffusion-based model, and the ground-truth(GT).}
    \label{fig:comparison_face_64}
\end{figure}

\paragraph{Results}
We trained our student model on the FFHQ training set using the pre-trained ResShift teacher model, which was originally trained on ImageNet. Although the teacher model was not specifically designed for human face data, our student model performs well on face data. Table~\ref{tab:FFHQ256} presents the quantitative results for 64$\times$64 images with $\times$4 SR, showing that our method outperforms other approaches in terms of PSNR, SSIM, and FID metrics. Qualitative comparisons of the LR input, our SR results, SinSR and the ground-truth are shown in Fig.~\ref{fig:comparison_face_64}.

\begin{table}[]
\centering
\vspace{4pt}
\begin{tabular}{c|cccc}
\hline
\multirow{2}{*}{Methods} & \multicolumn{4}{c}{Metric}                                        \\ \cline{2-5} 
                         & PSNR(↑)           & SSIM(↑)           & MS-SSIM(↑)        & FID(↓)            \\ \hline
SRGAN                    & 17.57          & 0.415          & 0.757          & 156.07         \\
ESRGAN                   & 15.43          & 0.267          & 0.747          & 166.36         \\
EnhanceNet               & 23.64          & 0.701          & 0.897          & 116.38         \\
SRFBN                    & 21.96          & 0.693          & 0.895          & 132.59         \\
CAGFace                  & 27.42          & 0.816          & \textbf{0.958} & 74.43          \\
SinSR(single step)       & {\underline {29.15}}    & {\underline {0.822}}    & 0.952          & {\underline {10.332}}   \\ \hline
Ours(single step)        & \textbf{29.43} & \textbf{0.829} & {\underline {0.955}}    & \textbf{9.179} \\ \hline
\end{tabular}
\caption{Quantitative results for 256x256 outputs on FFHQ dataset. The best and second best results of each metric are highlighted in \textbf{bold} and \underline{underlined}.}
\label{tab:FFHQ256}
\end{table}

\begin{table}[!htbp]
\centering
\begin{tabular}{ccc}
\hline
                     & Num of Iters & s/Iters \\ \hline
SinSR                & 30k          & 4.55s   \\
Ours (w Diff-based) & 20k          & 6.63s   \\
Ours (w Conv-based) & 20k          & 4.24s   \\ \hline
\end{tabular}
\caption{A comparison of the training cost measured on an NVIDIA  A100. Models using a diffusion-based downsampler are denoted as \textit{diff-based}, while those using a convolution-based downsampler are denoted as \textit{conv-based}.}
\label{tab:training_cost}
\end{table}

\subsection{Training overhead}
We compare the training cost of the proposed method with SinSR, another one-step diffusion model. As shown in Table ~\ref{tab:training_cost}, while SinSR achieves a shorter s/Iter than our method, which uses a diffusion downsampler, our model requires fewer iterations to converge. 

\subsection{Ablation Study}

We conducted an ablation study to evaluate the combination of three loss paths, as shown in Table~\ref{tab:ablation real}. The study compares the following losses: \textit{the distillation loss (Distill)} described in Sec~\ref{sec:3.1}, \textit{the HR loss} based on the student’s SR images with an HR discriminator, as detailed in Sec~\ref{sec:3.2}, and \textit{the LR loss} using a learnable downsampler with an LR discriminator, also explained in Sec~\ref{sec:3.3}. For this study, the downsampler was implemented using convolutional layers, while results marked with O* indicate experiments conducted with a diffusion-based architecture.
We evaluated performance on real-world datasets, RealSR and RealSet65, using the CLIPIQA and MUSIQ metrics. The results show that all three losses are crucial. While combining the distillation loss with either the HR loss or the LR loss yielded performance improvements, the best results were achieved when all three losses were combined.
Additionally, the diffusion-based approach results in notable performance gains. Importantly, our study is the first to introduce a diffusion-based downsampler that learns directly through a downsampling task while achieving upsampling as a result. This novel approach highlights the significance of our work in advancing the application of diffusion-based architectures.

\begin{table}[]
\centering
\vspace{4pt}
\resizebox{\columnwidth}{!}{%
\begin{tabular}{ccc|cccc}
\hline
\multicolumn{3}{c|}{\multirow{2}{*}{Loss}} & \multicolumn{4}{c}{Datasets}                                                                     \\ \cline{4-7} 
\multicolumn{3}{c|}{}                      & \multicolumn{2}{c|}{\textit{RealSR}}                    & \multicolumn{2}{c}{\textit{RealSet65}} \\ \hline
Distill          & HR         & LR         & CLIPIQA(↑)      & \multicolumn{1}{c|}{MUSIQ(↑)}         & CLIPIQA(↑)        & MUSIQ(↑)           \\ \hline
O*               & O*         & O*         & \textbf{0.7240} & \multicolumn{1}{c|}{\textbf{63.2628}} & \textbf{0.7428}   & 64.0630            \\
O                & O          & O          & {\underline {0.7030}}    & \multicolumn{1}{c|}{61.6313}          & {\underline {0.7329}}      & \textbf{64.5466}   \\
O                & O          & X          & 0.6909          & \multicolumn{1}{c|}{{\underline {62.0519}}}    & 0.7036            & {\underline {64.2709}}      \\
O                & X          & O          & 0.6887          & \multicolumn{1}{c|}{61.6081}          & 0.7076            & 63.3870            \\
O                & X          & X          & 0.6221          & \multicolumn{1}{c|}{58.7344}          & 0.6564            & 60.9495            \\ \hline
\end{tabular}%
}
\caption{Ablation Study on real-world datasets. Results marked with O* represent those obtained with the diffusion-based downsampler. The best and second best results are highlighted in \textbf{bold} and {\underline {underlined}}.}
\label{tab:ablation real}
\end{table}



%

\section{CONCLUSIONS}

In this work, we introduced a single-step diffusion model with two discriminators to enhance inference efficiency while maintaining high generative performance. Our method updates the student network cyclically, incorporating both HR and LR perspectives, and integrates adversarial loss via the discriminators. We also introduced a learnable diffusion-based downsampler to capture diverse degradation patterns. Even with synthetic LR-HR pairs, our approach generates multiple LR and HR samples at each iteration, leveraging degradation patterns to produce more realistic SR results. We demonstrated the effectiveness of our approach on the Real-ISR and Face-SR tasks. While it outperforms prior one-step methods, challenges remain in fine-scale details, such as small scene text. We anticipate that training on larger datasets could further improve the model’s overall generative capabilities.


\section*{ACKNOWLEDGMENT}
The authors would like to thank KT Corporation for providing GPU resources, which enabled this research.

{
\bibliographystyle{IEEEtran}
\bibliography{IEEEabrv, main} 
}

\end{document}